\documentclass[a4paper,11pt]{article}
\usepackage{pos}
\usepackage{subfigure}
\usepackage{lineno}
\usepackage{caption}
\renewcommand{\figurename}{Fig.}

\title{Measurement of transverse polarization of $\Lambda$/$\overline{\Lambda}$ within jet in $pp$ collisions at STAR}

\author*[a]{Taoya Gao}
\author[]{,For the STAR collaboration}

\affiliation[a]{Institute of Frontier and Interdisciplinary Science \& Key Laboratory of Particle Physics and Particle Irradiation(Ministry of Education),
  Shandong University, Qingdao, Shandong, 266237, China,}

\emailAdd{gao.ty@mail.sdu.edu.cn}

\abstract{Spontaneous polarization of $\Lambda/\overline{\Lambda}$ in unpolarized hadron interactions has been observed experimentally for nearly half a century and still eludes a definitive explanation. One possible origin is the effect arising from polarizing fragmentation functions (pFFs), which describe the production of polarized hadrons from the fragmentation of an unpolarized parton. Recently, significant transverse polarization of $\Lambda/\overline{\Lambda}$ has been observed in unpolarized $e^{+}e^{-}$ annihilation at Belle experiment, along the normal to the plane defined by the thrust axis and $\Lambda$ momentum. In unpolarized $pp$ collisions, the measurement of transverse polarization of $\Lambda$/$\overline{\Lambda}$ within jet could also provide important constraints and universality test for the pFFs. In this contribution, preliminary results on the first measurement of $\Lambda$/$\overline{\Lambda}$ polarization within a jet in $pp$ collision at $\sqrt{s}$ = 200 GeV are reported. The data used for this measurement were taken by the STAR experiment at RHIC in 2015.}

\FullConference{25th International Spin Physics Symposium (SPIN 2023)\\
 24-29 September 2023\\
 Durham, NC, USA\\}

\begin{document}
\captionsetup[figure]{labelfont={bf},labelformat={default},labelsep=period,name={Fig.}}
\maketitle

\section{Introduction}
The $\Lambda$ hyperon characterized by self-analyzing weak decay has played a special role in the field of spin physics \cite{Yang_1957}. In 1976, the large transverse polarization of hyperon was first observed in unpolarized $p+Be$ scattering \cite{Bunce_1976}, in a direction transverse to the production plane. Based on perturbative Quantum Chromodynamics (pQCD) calculations, the contributions from the hard scattering of hadronic collisions were found to be close to zero \cite{Kane_1978}. This discrepancy came as a surprise to the community. Numerous experiments followed to study the  $\Lambda$ spontaneous polarization in various reactions, $i.e.$, semi-inclusive deep inelastic scattering (SIDIS), $e^{+}e^{-}$ annihilation, hadron-hadron and hadron-nucleus scatterings \cite{Review_1990, HEAR, ATLAS, ALEPH, OPAL}. Despite lots of efforts and progress in understanding the $\Lambda$ polarization phenomenon, a definite explanation has not been identified.

One possible contribution could be from the Boer-Mulders function \cite{Boer_mulders} in the initial state, which describes the correlation between the transverse spin and intrinsic transverse momentum of quarks in an unpolarized nucleon. Another possible contribution could be from polarizing fragmentation functions (pFFs) \cite{pFFs,pFFs-pp} in the final state, which describe the production of a polarized hadron from the fragmentation of an unpolarized parton. In recent years, the study of polarizing fragmentation functions has received increasing attention as an effective tool to understand the fragmentation process \cite{TMD Alesio,TMD Kang, Liang_2021}, especially after the observation of the significant transverse polarization of $\Lambda(\overline{\Lambda})$ in $e^{+}e^{-}$ annihilation at Belle \cite{Belle_2019}. 

\begin{figure}[h]
\centering
{\includegraphics[width=0.6\linewidth]{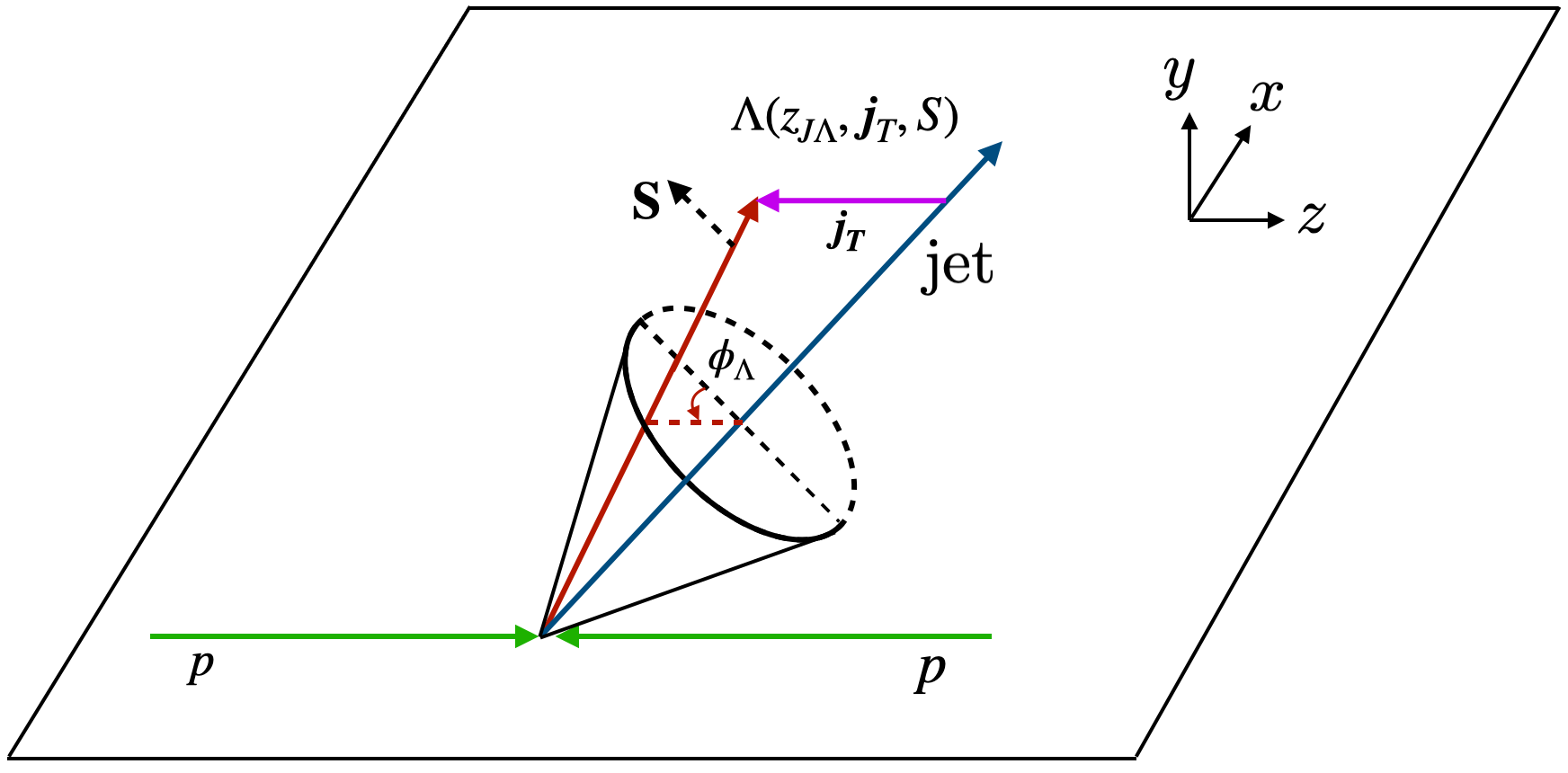}}
\caption{The illustration of $\Lambda$ hyperon production inside a jet in $pp$ collisions, vector $\textbf{S}$ denotes polarization direction defined by jet and $\Lambda$ momentum: $\textbf{S} = \textbf{p}_{jet} \times \textbf{p}_\Lambda$.}
\label{fig:jet_L}
\end{figure}

In $pp$ collisions, pFFs can be accessed by measuring polarization of $\Lambda$ within a jet \cite{jetPFF_2020}. And one advantage compared to a fixed energy scale in $e^{+}e^{-}$ annihilation is that a wide jet $p_T$ range can be obtained in $pp$ collisions. Therefore, we can study the energy scale dependence of pFFs by measuring the $\Lambda$ polarization versus jet $p_T$ in $pp$ collisions. Additionally, the process universality of pFFs could also be tested. The polarization direction of $\Lambda$ is defined along the normal direction to the plane defined by the jet and $\Lambda$ momenta as illustrated in \figurename{} \ref{fig:jet_L}, $\textbf{S} = \textbf{p}_{jet} \times \textbf{p}_\Lambda$.

\section{$\Lambda$ and jet reconstruction}
The $pp$ collision data at $\sqrt{s}$ = 200 GeV used for this measurement were collected by the STAR experiment at RHIC in 2015. The STAR detector comprises a variety of subdetectors. The Time Projection Chamber (TPC) \cite{TPC}, Barrel Electronmagnetic Calorimeter (BEMC) and Endcap Electronmagnetic Calorimeter (EEMC) \cite{BEMC,EEMC} are used in this analysis. The TPC provides charged-particle tracking and particle identification. The BEMC and EEMC are used for electromagnetic energy measurement and event triggering. In this analysis, only events triggered by JP1, one of the STAR jet-patch triggers with the threshold of 5.4 GeV, are used. 

The $\Lambda(\overline{\Lambda})$ candidates are reconstructed via the weak decay channel: $\Lambda \rightarrow p + \pi^{-}$ $(\overline{\Lambda}\rightarrow \overline{p} + \pi^+)$. A set of topological selection criteria is applied to pair two tracks with opposite charges to suppress the background, following similar procedure as in Ref. \cite{STAR_2024} except that the Time of Flight hit matching is not required for the pion track. The residual background from random track combinations and wrong particle identification is estimated to be about 10\% by using the side-band method and also subtracted. 

In order to measure the transverse $\Lambda$ polarization inside a jet, we need to reconstruct a jet including $\Lambda/\overline{\Lambda}$ particle. Such jet reconstruction is based on the TPC primary tracks, BEMC/EEMC tower energies and the reconstructed $\Lambda/\overline{\Lambda}$ \cite{STAR_2024}, using anti-$k_T$ algorithm with R = 0.6 and $p_T^{jet}$ > 5 GeV/$c$.  To suppress the edge effects,  jet $p_T$ is further required to be larger than 8 GeV/$c$.  To take into account the contributions from pile-up events or other background to jet reconstruction in $pp$ collisions, the off-axis method \cite{off-axis} is used to correct for the underlying event.

\section{Results}
The transverse polarization of $\Lambda$ is extracted via the angular distribution of the daughter particle in the $\Lambda$ rest frame:

\begin{equation}
    \frac{dN}{d\mathrm{cos}\theta^*}\propto A(\mathrm{cos}\theta^*)(1 + \alpha_{\Lambda(\overline{\Lambda})}P_{\Lambda(\overline{\Lambda})}\mathrm{cos}\theta^*),
    \label{eq:weak_decay}
\end{equation}
where $A(\mathrm{cos}\theta^*)$ is the acceptance function, $\theta^*$ is the angle between $\Lambda$ polarization direction and its daughter $p$ in the $\Lambda$ rest frame, $\alpha_{\Lambda/\overline{\Lambda}}$ = $\pm$ 0.732 is the decay parameter \cite{PDG_2020} and $P_{\Lambda(\overline{\Lambda})}$ is transverse polarization of $\Lambda$.

The angular distribution in Eq. (\ref{eq:weak_decay}) is sensitive to the detector acceptance which has to be corrected for. The detector acceptance function is estimated based on Monte-Carlo simulation by passing the $pp$ events generated by PYTHIA6.4.28 through GEANT3 framework of STAR detector. In addition, the same analysis procedure is applied to the MC sample as it is to the data. After acceptance correction, the polarization is extracted through fitting $\mathrm{cos}\theta^*$ distribution by a linear function. Limited by the size of the Monte-Carlo sample, the statistical uncertainties of the acceptance function dominate the uncertainties of the extracted transverse polarization.

\begin{figure}[h]
\centering
{\includegraphics[width=0.6\linewidth]{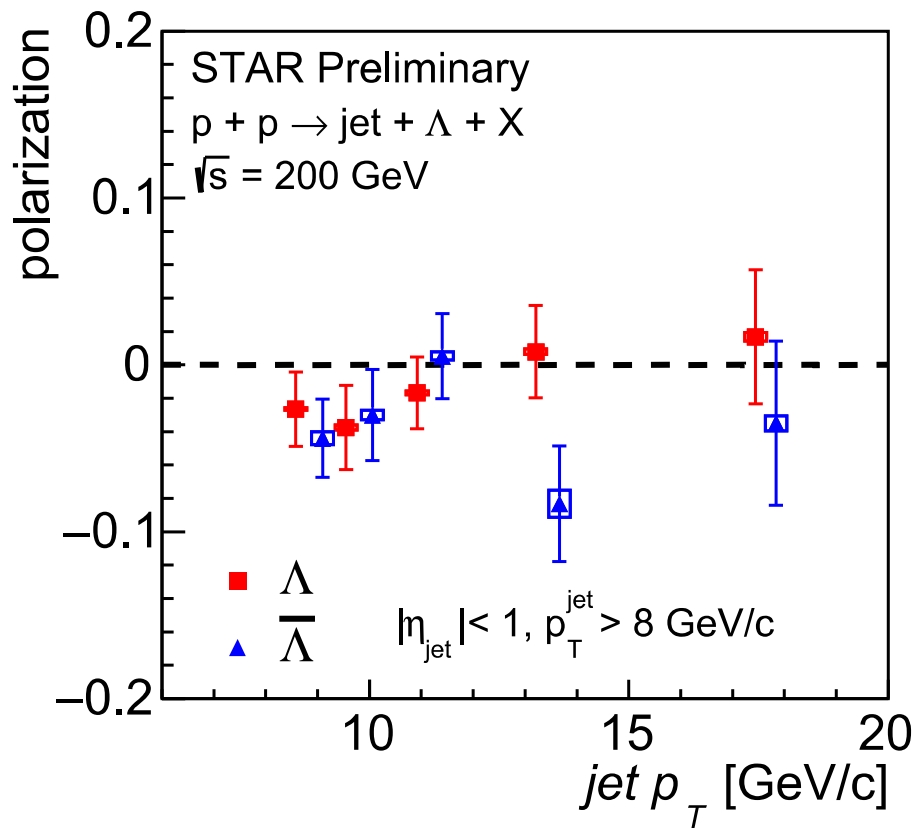}}
\caption{Preliminary results $\Lambda$ and $\overline{\Lambda}$ polarization within a jet versus jet $p_T$ in unpolarized $pp$ collisions at $\sqrt s=$ 200 GeV at STAR.}
\label{fig:Pol_jetPt}
\end{figure}

Figure \ref{fig:Pol_jetPt} shows the preliminary results on the transverse polarization of $\Lambda(\overline{\Lambda})$ versus jet $p_T$ in $pp$ collisions at $\sqrt s=$ 200 GeV. The average value of jet $p_T$ is about 11 GeV/$c$. Both $\Lambda$ and $\overline{\Lambda}$ indicate a hint of negative transverse polarization and also a weak dependence of jet $p_T$ at current precision. The magnitude of $\overline{\Lambda}$ polarization has a trend of increasing with jet $p_T$. This is the first hint of non-zero transverse polarization of $\Lambda(\overline{\Lambda})$ inside jet in unpolarized $pp$ collision.

\begin{figure}[h]
\centering
{\includegraphics[width=0.45\linewidth]{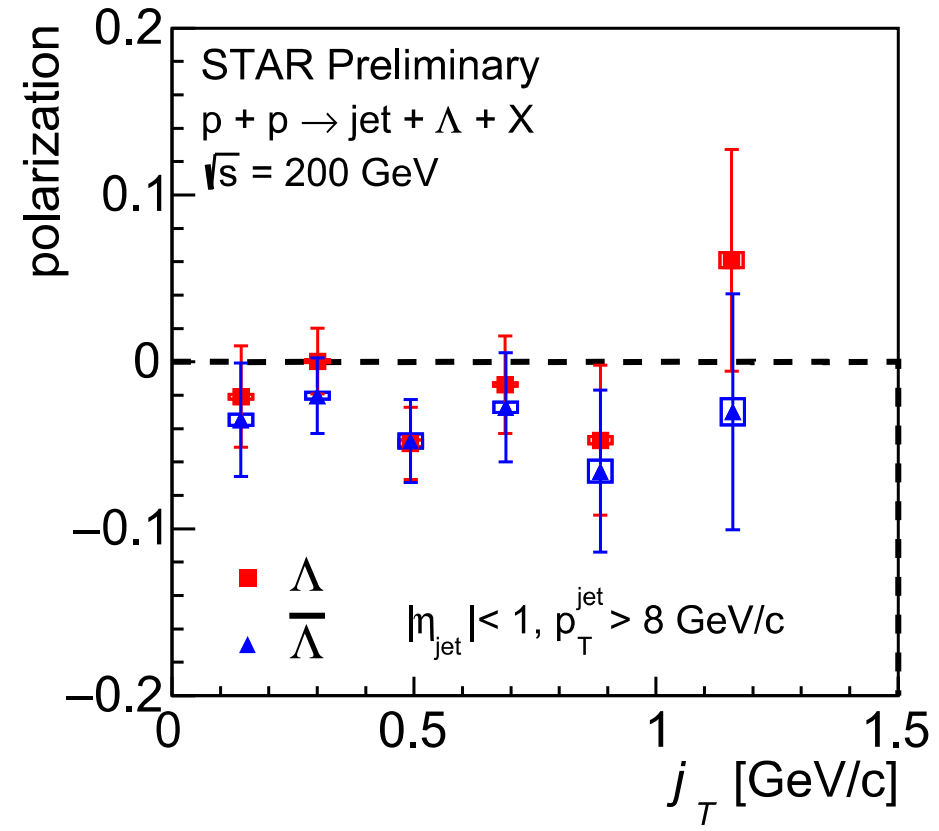}}
{\includegraphics[width=0.45\linewidth]{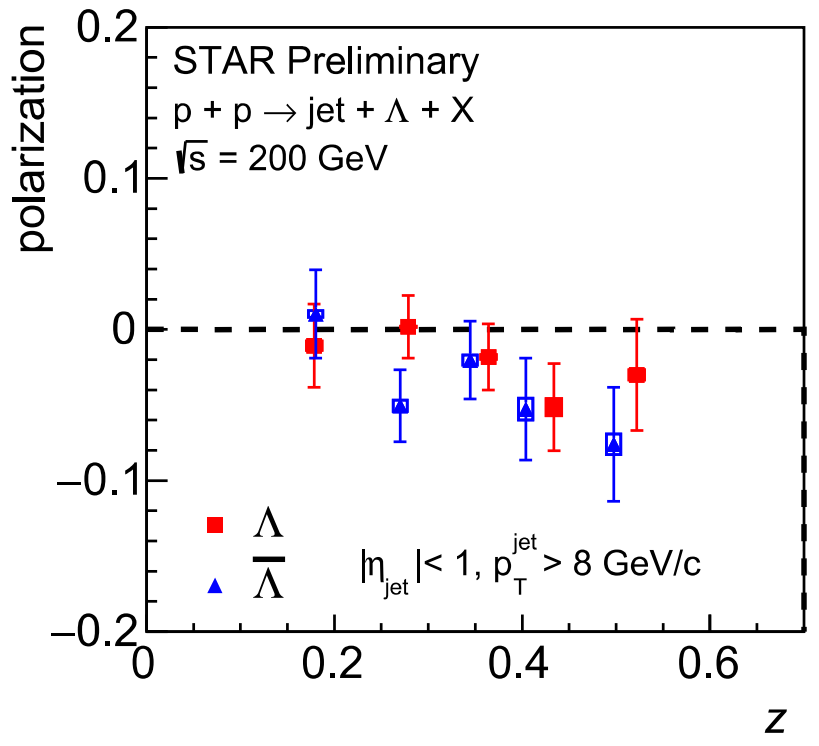}}
\caption{Preliminary results of $\Lambda$ and $\overline{\Lambda}$ polarization within a jet as a function of transverse momentum $j_T$ (Left), and jet momentum fraction $z$ (Right) in unpolarized $pp$ collisions at $\sqrt s=$ 200 GeV.}
\label{fig:Pol}
\end{figure}

To provide further constraints for the pFFs, the transverse polarizations of $\Lambda$ and $\overline{\Lambda}$ are also measured as functions of $j_T$ and $z$, as shown in \figurename{} \ref{fig:Pol}. As illustrated in Fig. 1, $j_T$ is the transverse momentum of $\Lambda$/$\overline{\Lambda}$ w.r.t. the jet axis, while $z$ is the momentum fraction of jet carried by $\Lambda$/$\overline{\Lambda}$. From the left panel of \figurename{} \ref{fig:Pol}, no $j_T$ dependence are observed for the transverse polarization of either $\Lambda$ or  $\overline{\Lambda}$. The right panel of \figurename{} \ref{fig:Pol} shows the transverse polarization as a function of $z$. There is a weak trend that the magnitude of the measured polarization increase with $z$ for both $\Lambda$ and $\overline{\Lambda}$.

\section{Summary}
The polarizing fragmentation functions (pFFs) is one of the most possible origins of the $\Lambda$ spontaneous polarization and can be accessed by measuring polarization of  $\Lambda(\overline{\Lambda})$ inside a jet in $pp$ collision at RHIC. In this contribution, we present the preliminary results on the first measurement of the transverse polarization of  $\Lambda(\overline{\Lambda})$ within a jet in $pp$ collision at $\sqrt{s}$ = 200 GeV. We observe a hint of negative polarization of $\Lambda(\overline{\Lambda})$ inside a jet, which also has a weak jet $p_T$ dependence with the current statistical precision. The transverse polarization as a function of $j_T$ and $z$ is also presented. These measurements could provide important constraints on pFFs like the scale evolution and the universality test. 
Improving the precision of the measurement, in particular the acceptance function is underway.  

\section*{Acknowledgements}
The author is supported partially by the National Natural Science Foundation of China under No. 12275159 and No. 12075140.


\begin{thebibliography}{99}
\bibitem{Yang_1957} T. D. Lee and C. N. Yang, Phys. Rev. \textbf{108}, 1645 (1957).
\bibitem{Bunce_1976}G. Bunce et al., Phys. Rev. Lett.  \textbf{36}, 1113 (1976).
\bibitem{Kane_1978}G. L. Kane, J. Pumplin, and W. Repko, Phys. Rev. Lett.  \textbf{41}, 1689 (1978).
\bibitem{Review_1990} A.D. Panagiotou, Int.J.Mod.Phys.A 5, 1197 (1990).
\bibitem{HEAR} I. Abt et al. [HERA-B collaboration], Phys. Lett. B \textbf{638}, 415 (2006).
\bibitem{ATLAS} G. Aad et al. [ATLAS Collaboration], Phys. Rev. D \textbf{91}, 032004 (2015).
\bibitem{ALEPH}D. Buskulic et al. [ALEPH collaboration], Phys. Lett. B \textbf{374}, 319(1996). 
\bibitem{OPAL}K. Ackerstaff et al. [OPAL Collaboration], Eur.Phys.J.C 2:49-59, (1998).
\bibitem{Boer_mulders}D. Boer and P. J. Mulders, Phys. Rev. D \textbf{57}, 5780 (1998).
\bibitem{pFFs}P. J. Mulders and R. D. Tangerman, Nucl. Phys. B \textbf{461}, 197 (1996).
\bibitem{pFFs-pp} M. Anselmino, D. Boer, U. D’Alesio, and F. Murgia, Phys.
Rev. D \textbf{63}, 054029 (2001).
\bibitem{TMD Alesio}U. D’Alesio, F. Murgia, and M. Zaccheddu, Phys. Rev. D \textbf{102}, 054001 (2020).
\bibitem{TMD Kang}D. Callos, Z.-B. Kang, and J. Terry, Phys. Rev. D \textbf{102}, 096007 (2020).
\bibitem{Liang_2021}K.B. Chen, Z.T. Liang, Y.L. Pan, Y.K. Song, and S.Y. Wei, Phys. Lett. B \textbf{816}, 136217 (2021).
\bibitem{Belle_2019}Y. Guan et al. [Belle collaboration], Phys. Rev. Lett.  \textbf{122}, 042001 (2019).
\bibitem{jetPFF_2020}Z.-B. Kang, K. Lee, F. Zhao,
Phys. Lett. B \textbf{809}, 135756 (2020). 
\bibitem{TPC} M. Anderson et al. [STAR Collaboration], Nucl. Instrum. Meth. A \textbf{499}, 659 (2003).
\bibitem{BEMC} M. Beddo et al. [STAR Collaboration], Nucl. Instrum. Meth. A \textbf{499}, 725 (2003).
\bibitem{EEMC} C.E. Allgower et al. [STAR Collaboration], Nucl. Instrum. Meth. A \textbf{499}, 740 (2003).
\bibitem{STAR_2024}M. I. Abdulhamid et al. [STAR collaboration], Phys. Rev. D \textbf{109}, 012004 (2024)
\bibitem{off-axis} B. Abelev et al. [ALICE Collaboration], Phys. Rev. D \textbf{91}, 112012 (2015).
\bibitem{PDG_2020} P.A. Zyla et al. (Particle Data Group), Prog. Theor. Exp. Phys. \textbf{2020}, 083C01 (2020).





\end{thebibliography}
\end{document}